# Public-private research collaborations: longitudinal field-level analysis of determinants, frequency and impact

Giovanni Abramo[1], Francesca Apponi[2] and Ciriaco Andrea D'Angelo[3]

[1] *giovanni.abramo@iasi.cnr.it*
Laboratory for Studies in Research Evaluation, Institute for System Analysis and Computer Science (IASI-CNR), National Research Council of Italy (Italy)

[2] *francesca.apponi@uniroma2.it*
Department of Engineering and Management, University of Rome "Tor Vergata" (Italy)

[3] *dangelo@dii.uniroma2.it*
Department of Engineering and Management, University of Rome "Tor Vergata" (Italy)
&
Laboratory for Studies in Research Evaluation, Institute for System Analysis and Computer Science (IASI-CNR), National Research Council of Italy (Italy)

**Abstract**

This study on public-private research collaboration measures the variation over time of the propensity of academics to collaborate with colleagues from private companies. It also investigates the change in weights of the main drivers underlying the academics' propensity to collaborate, and whether the type profile of the collaborating academics changes. To do this, the study applies an inferential model on a dataset of professors working in Italian universities in consecutive periods, 2010-2013 and 2014-2017. The results, obtained at overall and field levels, support the formulation of policies aimed at fostering public-private research collaborations, and should be taken into account in post-assessment of their effectiveness.

# 1. Introduction

Governments are increasingly turning to policy actions that can better enable industry to exploit the economic benefits of academic research (Fan, Yang, & Chen, 2015; Shane, 2004). One of the main modes of achieving the desired transfer of knowledge is through public-private research collaboration, and therefore a common policy aim is to increase the frequency and impact of collaboration. To achieve their objectives, policy-makers would be assisted by a better understanding of the motivations underlying joint cooperation.

The main objective of this work is therefore to investigate the changes over time in the frequency and impact of public-private research collaborations, alongside the change in weights of the main drivers underlying the academics' propensity to engage in such collaborations. With this knowledge, decision makers should be able to formulate incentive schemes that exploit the drivers with greater weight, and those that demonstrate greater long-term stability. A critical step in the study is the identification of the potential main drivers of the propensity of academics to engage in industry research collaborations. Previous literature suggests that the determinants of variety and frequency lie above all in the characteristics of the individual researcher, but to some extent also in the organizational or working environment, with further influences depending on their field of research (D'Este & Patel, 2007; Zhao, Broström, & Cai, 2020; Llopis, Sánchez-Barrioluengo, Olmos-Peñuela, & Castro-Martínez, 2018; Abramo, D'Angelo, & Murgia, 2013).

Previous studies investigating the drivers of academic engagement in collaboration with industry have been based on surveys, which embed severe limits on the scale of observations (Zhao, Broström, & Cai, 2020; Weerasinghe & Dedunu, 2020; Llopis, Sánchez-Barrioluengo, Olmos-Peñuela, & Castro-Martínez, 2018; Thune, Reymert, Gulbrandsen, & Aamodt, 2016). To overcome these, we adopt a bibliometric approach in which we analyze the co-authorships of publications: although there can be exceptions (Katz & Martin, 1997), co-authorship will in general be the outcome of a real research collaboration.

Our essential research questions are:
- Does the frequency and impact of public-private research collaboration vary over time?
- Does the typical profile of academics collaborating with the private sector change over time?
- Over time, is there change in the main individual and contextual drivers of propensity for private sector collaboration?
- Over time, how does the intensity and impact of collaboration vary across research fields?
- Over time, how does the importance of the main drivers vary across research fields?

To address these questions, we apply an inferential model on a dataset of professors working in Italian universities in two successive periods, 2010-2013 and 2014-2017. In general, when interpreting results of time-series analyses of the scientific output of research, in our case a joint research project, it must be kept in mind that output at time $t$ refers to a project conducted at time $t$-$\tau$, where $\tau$ represents the time lapse needed for the new knowledge to be encoded in written form and arrive at publication in a scientific journal. This issue of time lapse is especially important when designing systems intended to assess the effectiveness of policies fostering public-private research collaborations.

In the next two sections we review the literature on the factors determining private-public research collaboration, then present the methodology and data. In section four we show the results of the analysis at overall and at discipline level. Section five concludes the work.

# 2. Drivers of public-private research collaborations

The academic and their industrial research partner will have different motivations for engagement in research collaboration. The main aim of the company is to access knowledge with potential for commercial exploitation (Bekkers & Bodas Freitas, 2008; Perkmann et al., 2013). The academic is primarily attracted by the company's offer of direct economic and financial benefits (Garcia, Araújo, Mascarini, Santos, & Costa, 2020), and to some extent, by a noble purpose: the so called "mission" motivation, i.e. fulfillment of the university's societal role (Iorio, Labory, & Rentocchini, 2016). This means that the intensity of public-private research collaboration would be sensitive over time to different personal and contextual issue, among these: the academics' changing financial needs; the adoption of incentive schemes by national and regional governments, as well as by their employer organizations; the establishment and effectiveness of university-industry liaison offices as catalysts for collaboration.

For the academics, propensity to collaborate with industry varies along the career cycle. Collaboration propensity appears stronger at the early career stage and weaker in the later stages (Bozeman & Gaughan, 2011; Ubfal & Maffioli, 2011), although the impact of age seems non-linear (Weerasinghe & Dedunu, 2020). Younger academics are in more need of financial resources but will also aim to master the activation and management of collaborations, in part for benefits of positive evaluation and career progress (Bayer & Smart, 1991; Traoré & Landry, 1997).

Because of gender homophily, women experience greater challenges in forming social capital, including in developing collaboration networks (Boschini & Siogren, 2007). Female academics also seem less attracted to industrial collaboration (Tartari & Salter, 2015), and would therefore engage less, (Calvo, Fernández-López, & Rodeiro-Pazos, 2019).

Private collaboration correlates positively with individual research performance (Mansfield, 1995; Mansfield & Lee, 1996; He, Geng, & Campbell-Hunt, 2009; Lee & Bozeman, 2005; Schartinger, Schibany, & Gassler, 2001), since companies searching for research partners are more likely to engage with top players (Balconi, & Laboranti, 2006).

Academics who tend to diversify their research activities could have underlying characteristics of flexibility and curiosity for the unexplored, leading them to be more likely to engage beyond the academic sphere (Abramo, D'Angelo, & Di Costa, 2019).

The academic's propensity for industrial collaboration would also be influenced by environmental factors, in particular the importance placed by their home university on technology transfer. From this there descends an institutional culture with intrinsic motivational stimuli, that are the harbingers of research collaboration with industry (Di Gregorio & Shane, 2003; Giuri, Munari, Scandura, & Toschi, 2019). At the institutional level and beyond, there would also be the question of the availability of financial resources, in turn affecting the individual's propensity to search for further, private sources (Giuri, Munari, Scandura, & Toschi, 2019).

The inclination and attitude of the academic's colleagues towards cross-sector research projects, operating through mechanisms of social comparison, will influence their own behavior (Tartari, Perkmann, & Salter, 2014). The university's size can influence faculty in choosing different types of collaboration, for example when below a threshold of critical research mass, the academics would likely engage more in extramural collaboration (Schartinger, Schibany, & Gassler, 2001).

Considering universities within a given territory, the relative concentration of industrial R&D will affect the demand for research collaboration, and therefore the possibilities for academic-industry collaboration (Berbegal-Mirabent, Sánchez García, & Ribeiro-Soriano, 2015).

Finally, the intensity of publication is known to vary across research fields (D'Angelo & Abramo, 2015). Since we investigate research collaborations through their publication outputs, we should be able to detect these field effects, and for this purpose we will carry out field-level analysis.

## 3. Data and methods

The field of observation consists of professors of Italian universities conducting research in the so-called hard sciences. We exclude the social sciences, arts, and humanities, because for these the coverage of bibliographic repertories is still insufficient for reliable representation of research output. In the Italian university system all professors are classified in one and only one field (named the scientific disciplinary sector, SDS, 370 in all). Fields are grouped into disciplines (named university disciplinary areas, UDAs, 14 in all). We observe research production in 201 SDSs falling in 10 UDAs, where publications in international journals serve as a reliable proxy for overall research output.

To investigate whether the collaboration rates, their relevant impact, and the underlying determinants of academic engagement in joint research with industry change over time, we conduct a longitudinal analysis dividing the observed period into two subsequent four-year subperiods, 2010-2013 and 2014-2017.[1]

The dataset for the analysis was extracted from the Italian Ministry of University and Research (MIUR) database of Italian professors[2] and consists of all assistant, associate and full professors, on staff for at least three years in 2010-2013 or 2014-2017, with at least one Web of Science (WoS) indexed publication in the relevant period. Their distribution by UDA and academic rank is shown in Table 1.

*Table 1. Dataset of the analysis, by academic rank and discipline (UDA)*

| | 2010-2013 | | | | 2014-2017 | | | |
|---|---|---|---|---|---|---|---|---|
| UDA | Assistant | Associate | Full | Total | Assistant | Associate | Full | Total |
| 1 - Mathematics and computer science | 1167 | 748 | 810 | 2725 | 1226 | 751 | 707 | 2684 |
| 2 – Physics | 876 | 696 | 499 | 2071 | 973 | 642 | 420 | 2035 |
| 3 - Chemistry | 1333 | 844 | 607 | 2784 | 1405 | 790 | 509 | 2704 |
| 4 - Earth sciences | 444 | 315 | 217 | 976 | 482 | 288 | 181 | 951 |
| 5 – Biology | 2238 | 1196 | 1082 | 4516 | 2410 | 1121 | 881 | 4412 |
| 6 – Medicine | 4392 | 2520 | 1936 | 8848 | 4368 | 2359 | 1645 | 8372 |
| 7 - Agricultural and veterinary sciences | 1262 | 751 | 680 | 2693 | 1337 | 782 | 608 | 2727 |
| 8 - Civil engineering | 551 | 361 | 376 | 1288 | 622 | 391 | 337 | 1350 |
| 9 - Industrial and information engineering | 2059 | 1264 | 1336 | 4659 | 2269 | 1399 | 1268 | 4936 |
| 10 – Psychology | 535 | 280 | 268 | 1083 | 665 | 299 | 257 | 1221 |
| Total | 14857 | 8975 | 7811 | 31643 | 15757 | 8822 | 6813 | 31392 |

The bibliometric dataset is extracted from the Italian Observatory of Public Research (OPR), a database developed and maintained by the present authors, derived under license from the Thomson Reuters WoS. Beginning from the raw data of the WoS and applying a complex algorithm to reconcile the author's affiliation and disambiguation of their true identity, each publication (article, article review and conference proceeding) is attributed to the professors that authored it (D'Angelo, Giuffrida, & Abramo, 2011).[3] Collaboration with industry is evidenced by the presence of at least one private company in the address list of publications authored by the professors in the dataset. Such identification requires manual scrutiny and reconciliation of all bibliographic addresses with "Italy" as affiliation country.

---

[1] For observation of publication impact, we choose the end date of 2017 rather than a more recent one, considering that the longer the citation window the more robust and precise will be the bibliometric measurement. Furthermore, longer observation period would be costly, because of the time-consuming task of manually identifying and reconciling all bibliographic names and university/company addresses for the publications.

[2] For each individual, the MIUR database shows the last name and given names, university, SDS, academic rank and department. For details see http://cercauniversita.cineca.it, last access 31 January 2021.

The following publication provides an example:

> Ferrante, I., Ciprianetti, N., Stassi, S., Santoro, K., Ferrero, S., Scaltrito, L., & Ricciardi, C. (2017). High-throughput characterization of microcantilever resonator arrays for low-concentration detection of small molecules. *Journal of Microelectromechanical Systems, 26*(1), 246-254. DOI:10.1109/JMEMS.2016.2633315

The address list shows five records:
- Politecn Torino, Appl Sci & Technol Dept, I-10129 Turin, Italy
- Genewave SAS, Mobidiag France, 172 Rue Charonne, F-75011 Paris, France
- Trustech SRL, I-10034 Chivasso, Italy
- Univ Turin, Dept Agr Forest & Food Sci, I-10095 Grugliasco, Italy
- Microla Optoelectron, I-10034 Chivasso, Italy

Analysis of the four affiliations showing Italy as country reveals the presence of two universities (Politecnico di Torino, Università degli Studi di Torino) and two Italian companies (Trustech Srl, Microla Optoelectronics Srl). The initial stages of this type of analysis are carried out by applying rules of reconciliation developed by the authors and continuously updated since setting up the OPR. The analysis is carried out on about 700,000 publications with at least country of affiliation "Italy", indexed in WoS over 2010-2017.

While public-private research collaboration pursues scientific and technological advances, part of any new knowledge produced will remain tacit, not encoded in written forms such as publications and patents, making assessment of the overall impact of public-private collaborations a truly daunting task. In our case we concentrate on measuring the scholarly impact of collaboration outputs, meaning that we apply citation-based metrics to the knowledge encoded in the publications. We refer the reader to Abramo (2018) for a thorough discussion on the conceptualization of publication impact and its bibliometric measurement.

We count the citations to the roughly 700,000 publications (2010-2013; 2014-2017) as of 30/06/2019. This is still a relative short time citation window, so the citation counts alone will not guarantee strong prediction of the long-term impacts. To improve predictive strength, as proposed by Abramo, D'Angelo, and Felici (2019), we rely on a weighted combination - different for each subject category and year of publication - of the field-normalized citations and the field-normalized impact factor of the hosting journal (measured at year of publication).[4]

In the following, we first verify whether the share of professors collaborating with industry in the two periods increases or not, and if there is change over time in the type profile of those engaging in the collaborations. Then, we analyze whether there is change over time in the impact of the publications from the collaborations. Finally, we use a logit regression in order to understand if, between the two periods, there is any change in the relative weight of the main

---

[3] The harmonic average of precision and recall (F-measure) of authorships, as disambiguated by the algorithm, is around 97% (2% margin of error, 98% confidence interval).

[4] Because citation behaviour varies across fields, and the citations accumulated are also a function of time, we normalize citations by scaling the citation counts with the average of the distribution of citations received for all cited publications of the same year and subject category, previously identified as the most effective among possible scaling factors (Abramo, Cicero, & D'Angelo, 2012).

drivers of academic engagement in private collaboration. The logit regression relies on a dummy dependent variable (Y) assuming: 1, if professor *i* co-authored at least one publication with industry; 0, otherwise. Coherent with previous literature on covariates likely to affect the propensity of academics to engage in industry collaboration, we consider the following, listed in clusters of "individual" and "contextual":

Individual covariates
- Gender ($X_1$), specified by a dummy variable (1 for female; 0 for male);
- Age ($X_{2-5}$), specified with 5 classes, through 4 dummies (baseline "Less than 40");
- Academic rank ($X_{6-7}$), specified by 2 dummies (baseline "Assistant professors");
- Productivity - Total publications authored by the professor in the period under observation ($X_8$);
- Research type – The professor's level of specialization in production, specified by 2 dummies:[5]
    - "Highly diversified" ($X_9$), 1 if the papers falling in the prevalent subject category of the professor are less than 40% of total publications; 0, otherwise;
    - "Highly specialized" ($X_{10}$), 1 if the papers falling in the prevalent subject category of the professor are more than 75% of total publications; 0, otherwise;[6]

Contextual covariates
- University culture – Collaborating peers ($X_{11}$), specified by a dummy variable (1 in case of a colleague in the same university and SDS of the professor, co-authoring publications with industry; 0, otherwise);
- University type – Legal status ($X_{12}$), specified by a dummy variable (0, for public universities; 1, for private ones);
- University type - Scope ($X_{13}$), specified by a dummy variable (1, for "Polytechnics" and "Special Schools for Advanced Studies, SS"; 0, otherwise);[7]
- University size in the UDA of the professor ($X_{14-15}$), specified with 3 classes through 2 dummies ("Large", for universities with a research staff in the UDA of the professor, above 80 percentile in the national ranking; "Medium", with a research staff between 50 and 80 percentile);
- University location ($X_{16-19}$), specified with 5 geographical macro-areas, by 4 dummies (baseline "Islands").

To control for the effects of field of research, we also consider 9 dummies indicative of the ten UDAs under observation.

The logit regression was applied to the two periods under observation, with variables measured at 31/12/2009 for the first, 31/12/2013 for the second.

## 4. Results

---

[5] For a thorough explanation of the bibliometric approach to distinguishing specialized from diversified research, see Abramo, D'Angelo, and Di Costa (2017).
[6] The thresholds chosen (40% for $X_8$, and 75% for $X_9$) distinguish the dataset in equal parts: highly specialized and highly diversifed professors are one-third each.
[7] The Italian Minister of University and Research (MUR) recognizes 96 universities as having the authority to issue degrees. Of these, 29 are small, private, special-focus universities, of which 13 offer only e-learning, 67 are public and generally multi-disciplinary universities. Three of them are Polytechnics and six are *Scuole Superiori* (Special Schools for Advanced Studies), devoted to highly talented students, with very small faculties and tightly limited enrolment. In the overall system, 94.9% of faculty are employed in public universities, 0.5% in *Scuole Superiori*.

## 4.1 Analysis at overall level

In the 2010-2013 dataset, 17.2 percent of professors coauthored at least one publication with industry (Table 2, subset A); in the second dataset this share rose to 25.5 percent (subset B). The professors on staff across both 2010-2013 and 2014-2017 were 27,323; of these, 10.0 percent collaborated with industry in both periods, 8.1 percent in the first period only, and 16.0 percent in the second period only (i.e. combining these subsets, the total set collaborating in the first period was 18.1 percent, the total set in the second period amounted to 26.0 percent).

Two other subsets considered in Table 2 consist of the professors who left the academia in the second period (A - B) and joined it in the same period (B - A). Among the former, the share of those who co-authored at least one publication with industry in 2010-2013 was 11.3 percent. Among the latter, the share almost doubled (21.6%).

We can say then, that i) the share of professors on staff in both periods, collaborating with industry increased by 7.9 percentage points; and ii) the new entries (in the second period) showed a propensity to collaborate with industry almost double that of the professors departing, generally for retirement.[8]

*Table 2. Subsets of professors in the dataset*

| Subset* | No. | Category | Share |
|---|---|---|---|
| A | 31643 | Collaborating with industry | 17.2% |
| B | 31392 | Collaborating with industry | 25.5% |
| A ∩ B | 27323 | Collaborating in both periods | 10.0% |
| | | Collaborating only in the second period | 16.0% |
| | | Collaborating only in the first period | 8.1% |
| | | Never collaborating | 65.8% |
| A - B | 4320 | Collaborating with industry | 11.3% |
| B - A | 4069 | Collaborating with industry | 21.6% |

* "A" = professors in the 2010-2013 dataset; "B" = professors in the 2014-2017 dataset

The 2010-2013 publications co-authored with industry (i.e. by subset A) show an average field-normalized impact of 0.998 (Table 3). Those of 2014-2017 (i.e. by subset B) show a noticeably lower average impact (0.894). In both cases, average impact is below average, meaning that professors produce higher impact publications when not partnering with industry. Considering that professors' research performance is increasingly under scrutiny, this could lead to counter-purposes between policies aimed at stimulating researcher productivity and, contrastingly, at encouraging their engagement in public-private collaboration. For professors and university departments, the result would be unwelcome compromises and trade-offs.

The decrease in average impact between the two periods is due mainly to the collaborating professors who continue to work over both periods (A ∩ B), whose average impact from industry co-authored publications was 1.013 in 2010-2013, declining to 0.885 in 2014-2017. This is further confirmed if we look at the co-authored publications by collaborating professors who left academia over the second period (A - B), who show very low average impact (0.813), much lower than the impact (0.974) from professors arriving new in academia (B - A) and taking up collaboration. We deduce that those professors who choose to continue collaboration with industry over the long term will then experience negative effects on the impact of their production.

*Table 3. Average impact of publications from academic-industry collaboration*

---

[8] Almost all professors recorded in the first period but absent in the second had retired due to age limits. It is rare that any faculty leave an Italian university for reasons other than retirement (Abramo, D'Angelo, & Rosati, 2016).

| Subset* | 2010-2013 | 2014-2017 |
|---|---|---|
| A | 0.998 | - |
| B | - | 0.894 |
| A ∩ B | 1.013 | 0.885 |
| A - B | 0.813 | - |
| B - A | - | 0.974 |

* "A" = professors in the 2010-2013 subset; "B" = professors in the 2014-2017 subset

Table 4 presents the typical profile of university professors collaborating with private companies, detected by the concentration index on each individual and contextual covariate.[9] The left and right sides show the profiles emerging from the data of the first and second four-year periods.

In 2010-2013, the type academic active in industry collaboration is male, 40-45 years old, a full professor, conducting research in Industrial and information engineering, with highly diversified research activity. This professor operates within a group of peers who also engage with industry (i.e. collaborate), and belongs to a medium-sized public university, typically a polytechnic or SS located in northwestern Italy.

The 2014-2017 profile is very similar, differing significantly only in the type academic's age and the size of their home university. Regarding age, the prevailing trait descends from age 40-45 years to below 40 years: i.e. in the second period, younger academics assume a greater role in interaction with industry. The shift in the typical size of home institution is from medium to large, although the association of such variable to the academic propensity to collaborate with industry is not statistically significant in either period. The other type traits remain, although there is some variation in the absolute value of the concentration indices, revealing a weakening/strengthening of the trait.

Since this purely descriptive analysis does not inquire into the effects of all the covariates simultaneously acting on the independent variable, we conduct an inferential analysis using a logit regression model (illustrated in Section 2). Table 5 shows the average values of the model variables, at overall level. Some significant differences can be noted between the two periods, for individual covariates:

- The average age of the collaborating professors increases, as seen in the rising incidence of the older cohorts (e.g. the share of over-sixties increases from 12.6 percent to 14.3 percent of total).
- The collaborating professors show an increasing average number of publications, from 14.60 to 19.37.
- Among those collaborating with industry, the share of academics conducting highly diversified research increases, rising from 14.3 percent to 19.1 percent. Vice versa the share of specialized researchers decreases, from 31.3 percent to 25.2 percent.

As noted, the datasets of the two periods overlap, with about 87 percent of professors continuing through both. The variations over time would thus be partly explained by incumbent professors experiencing behavioral changes, and partly by new hires arriving with different attitudes and interests compared to retiring ones. In fact, over time, in spite of increasing average age (noted above), the share of associate professors, and even more so full professors, dropped in favor of assistant professors. Also notable is that gender balance among collaborating professors improved, with the mix of entries and exits resulting in the share of female professors rising from 32.3 to 33.9 percent. In fact, in the incoming subset "B - A" of Table 2, the share of females is 37.6 percent, while in outgoing subset "A - B" it is 25.3 percent.

---

[9] The concentration index is the ratio of two ratios. Example: for the group variable "gender", the prevailing trait "male" shows a concentration index of 1.058 for 2010-2013 data, since males compose 71.66% of total researchers co-authoring publications with industry, and 67.75% of the total population, therefore 71.66/67.75=1.058.

*Table 4: Type profiling Italian academics who co-author publications with industry*

|  | 2010-2013 | | | 2014-2017 | | |
|---|---|---|---|---|---|---|
| Variable | Prevailing trait | Concentrat. index | Pearson chi$^2$ | Prevailing trait | Concentrat. index | Pearson chi$^2$ |
| Gender | Male | 1.058 | 46.1*** | Male | 1.076 | 120.0*** |
| Age | 40-45 years | 1.073 | 54.9*** | Below 40 years old | 1.132 | 132.1*** |
| Academic rank | Full professor | 1.160 | 78.2*** | Full professor | 1.150 | 80.9*** |
| UDA | 9 - Ind.+Infor. Engineering | 1.661 | ≈1000*** | 9 - Ind.+Infor. Engineering | 1.777 | 1.9e+03*** |
| Scientific activity | Highly diversified | 1.328 | 356.5*** | Highly diversified | 1.220 | 306.3*** |
| Environment | Peers collaborating | 1.432 | ≈16000*** | Peers collaborating | 1.265 | 1.5e+03*** |
| University type | Public | 1.005 | 4.6** | Public | 1.012 | 35.6*** |
|  | Polytechnic or SS | 1.619 | 152.3*** | Polytechnic or SS | 1.676 | 309.0*** |
| University size | Medium | 1.014 | 0.7 | Large | 1.005 | 1.9 |
| University location | Northwest | 1.262 | 167.2*** | North-west | 1.148 | 149.1*** |

*Statistical significance: \*p-value <0.10, \*\*p-value <0.05, \*\*\*p-value <0.01*

*Table 5: Average values of the regression model variables*

| Variable | | | 2010-2013 | 2014-2017 |
|---|---|---|---|---|
| Response | Y | Co-authorships with industry | 0.172 | 0.255 |
| Gender | $X_1$ | Female | 0.323 | 0.339 |
| Age | $X_2$ | 40-45 | 0.200 | 0.190 |
|  | $X_3$ | 46-52 | 0.244 | 0.252 |
|  | $X_4$ | 53-60 | 0.229 | 0.258 |
|  | $X_5$ | Over 60 | 0.126 | 0.143 |
| Academic rank | $X_6$ | Associate professor | 0.284 | 0.281 |
|  | $X_7$ | Full professor | 0.247 | 0.217 |
| Productivity | $X_8$ | Total publications | 14.60 | 18.37 |
| Scientific activity | $X_9$ | Highly diversified | 0.143 | 0.191 |
|  | $X_{10}$ | Highly specialized | 0.313 | 0.252 |
| Environment | $X_{11}$ | Peers collaborating | 0.561 | 0.660 |
| University type | $X_{12}$ | Private | 0.034 | 0.039 |
|  | $X_{13}$ | Polytechnic or SS | 0.057 | 0.059 |
| University size | $X_{14}$ | Medium | 0.329 | 0.321 |
|  | $X_{15}$ | Large | 0.598 | 0.610 |
| University location | $X_{16}$ | South | 0.200 | 0.200 |
|  | $X_{17}$ | Center | 0.254 | 0.250 |
|  | $X_{18}$ | Northeast | 0.197 | 0.196 |
|  | $X_{19}$ | Northwest | 0.239 | 0.244 |

Table 6 presents the results of the logit regressions investigating the drivers of academic engagement in public-private research collaboration in the two periods. The UDA dummies were included in the modeling for control of effects related to scientific field, but are omitted from the table for simplicity of presentation.

The model estimation appears satisfactory. For the 2014-2017 period, the mean VIF is 2.49 (maximum 7.72 for covariate $X_{15}$, University size – Large), indicating the absence of significant multicollinearity. The AUC (area under ROC curve) is 0.737, indicating good ability to correctly discriminate the propensity of professors to collaborate with companies[10] in function of the individual and contextual covariables. The estimated coefficients are expressed in terms of odds ratios: the reference value is equal to one and indicates that the independent variable considered has no effect on the dependent variable, i.e. on the probability that a professor has or has not collaborated with private companies. For values above one, the variable instead has a positive marginal effect, and vice versa.

From Table 6, the data indicate that almost all the covariates have statistically significant effect. The variable $X_{13}$, University scope, is the only one without significant marginal effect over both periods, on propensity to collaborate with industry.

Confirming previous literature on the effect of individual traits (Weerasinghe & Dedunu, 2020; Tartari & Salter, 2015; Calvo, Fernández-López, & Rodeiro-Pazos, 2019), gender has a non-marginal effect: in the first period, compared to men, women show a lower propensity to collaborate with industry (-10.2 percent), still lower in the second (-13.0 percent). Age shows a systematically negative impact on the response variable, invariable over time: all odds ratios below 1 and decreasing similarly with age in both periods.

Academic rank shows positive effects on propensity for industry collaboration, decreasing slightly in the second period. Full and associate professors show greater disposition to collaborate compared to assistant professors, by 86 percent and 39 percent in 2010-2013, vs 77 percent and 33 percent in 2014-2017.

The analysis reveals a marginal effect of the professors' research diversification/specialization, increasing over time. In 2010-2013, a highly diversified scientific profile had a probability of collaborating with companies 32 percent higher than the "intermediate" profile, with effect increasing to +37% in 2014-2017. In contrast, the dual covariant (profile "highly specialized") has negative impacts on propensity to collaborate with industry.

Finally, the effect of what can be considered an individual exposure variable (total number of publications), although statistically significant, is very limited in both periods (odds ratio 1.012 in 2010-2013, and 1.010 in 2014-2017).

Among all covariates under examination, "Environment: peers collaborating" ($X_{11}$) presents the highest odds ratio, confirming the indication of Tartari, Perkmann, and Salter (2014) that the presence of colleagues in the same field, actively collaborating with companies, is the most important driver for an academic to similarly engage. Interestingly, however, the marginal effect in the Italian cases decreases substantially over 2010-2017 (from 3.472 odds ratio in the first period to 2.834 in the subsequent). Among other contextual covariates, the "public" aspect of a university has a positive effect on faculty propensity to collaborate, probably because the lesser financial resources compared to private universities encourage outreach to industry. Furthermore, all others equal, the faculty of large universities are less likely to collaborate with industry than faculty of small universities (-33.2 percent in the first period, -20.5 percent in the second). The coefficients of the four dummies for geographic localization indicate an increase in the propensity to collaborate with latitude (i.e. towards northern Italy, which has the highest concentration of private R&D), however the indications became more pronounced over time, since in the first period only the dummy "Northwest" showed statistically significant effect. Over time, the case of Italian university professors seems to conform increasingly with theoretical expectations: that interaction with companies depends on the level of concentration of industrial activities in general, and in particular of knowledge-intensive industry.

---

[10] The AUC analysis evaluates the ability of a receiver (receiver operating characteristic, RUC) to discern between true and false positives. In our case, the AUC value, between 0 and 1, is equivalent to the probability that the result of the logit classifier applied to a researcher randomly extracted from the group of those who collaborated with industry is higher than that obtained by applying it to a researcher randomly extracted from the group of those who did not collaborate (Bowyer, Kranenburg, & Dougherty, 2001).

*Table 6: Italian professors: Main drivers of the propensity to collaborate with industry. ( Logit regression, dependent variable: 1 in case of publications in co-authorship with industry, 0, otherwise.)*

| Variable | | | 2010-2013 | 2014-2017 |
|---|---|---|---|---|
| | | Const. | 0.040*** | 0.061*** |
| Gender | $X_1$ | Female | 0.898*** | 0.870*** |
| | $X_2$ | 40-45 | 0.916* | 0.960 |
| Age | $X_3$ | 46-52 | 0.787*** | 0.768*** |
| | $X_4$ | 53-60 | 0.617*** | 0.626*** |
| | $X_5$ | Over 60 | 0.449*** | 0.470*** |
| Academic rank | $X_6$ | Associate professor | 1.392*** | 1.332*** |
| | $X_7$ | Full professor | 1.862*** | 1.775*** |
| Productivity | $X_8$ | Total publications | 1.012*** | 1.010*** |
| Scientific activity | $X_9$ | Highly diversified | 1.320*** | 1.369*** |
| | $X_{10}$ | Highly specialized | 0.611*** | 0.631*** |
| Environment | $X_{11}$ | Peers collaborating | 3.472*** | 2.834*** |
| University type | $X_{12}$ | Private | 0.841* | 0.825** |
| | $X_{13}$ | Polytechnic or SS | 0.918 | 1.062 |
| University size | $X_{14}$ | Medium | 0.831*** | 0.892* |
| | $X_{15}$ | Large | 0.668*** | 0.795*** |
| University location | $X_{16}$ | South | 0.905 | 1.208*** |
| | $X_{17}$ | Center | 1.008 | 1.352*** |
| | $X_{18}$ | Northeast | 1.094 | 1.397*** |
| | $X_{19}$ | Northwest | 1.337*** | 1.464*** |
| | | Number of obs | 31643 | 31392 |
| | | LR chi2(28) | 3394.67 | 4189.84 |
| | | Prob > chi2 | 0.000 | 0.000 |
| | | Log likelihood | -12589.9 | -15661.6 |
| | | Pseudo $R^2$ | 0.119 | 0.118 |

*Statistical significance: *p-value <0.10, **p-value <0.05, ***p-value <0.01*

## 4.2 Analysis at discipline level

In the previous section, at the overall level from one four-year period to the next, we observed an increase in the rate of collaboration between academics and industry. Examining at the level of individual disciplines (Figure 1) this observation is substantially generalized. Across the two periods, the discipline with the highest percentage of professors involved in private-company collaborations is Industrial and information engineering. This discipline also shows the greatest absolute increase over the two periods, at over 16 percentage points: from 28.6%, of professors collaborating in the first period, to 45.2% in the second. UDAs 4 (Earth sciences) and 8 (Civil engineering) also record increases greater than 10 percentage points. Mathematics and computer science (UDA 1) and Psychology (UDA 10), were "tail-enders" in the first period and remain so in the second. In UDA 1, however, the incidence of public-private collaboration doubles, reaching 10.5% of total professors in the second period against 5.3% in the first. In both Medicine, the largest discipline by number, and Biology, professors showed only limited increase in collaboration with industry (+5.2% and +5.8%).

*Figure 1. Share of Italian professors co-authoring at least one publication with industry, by UDA and time period*

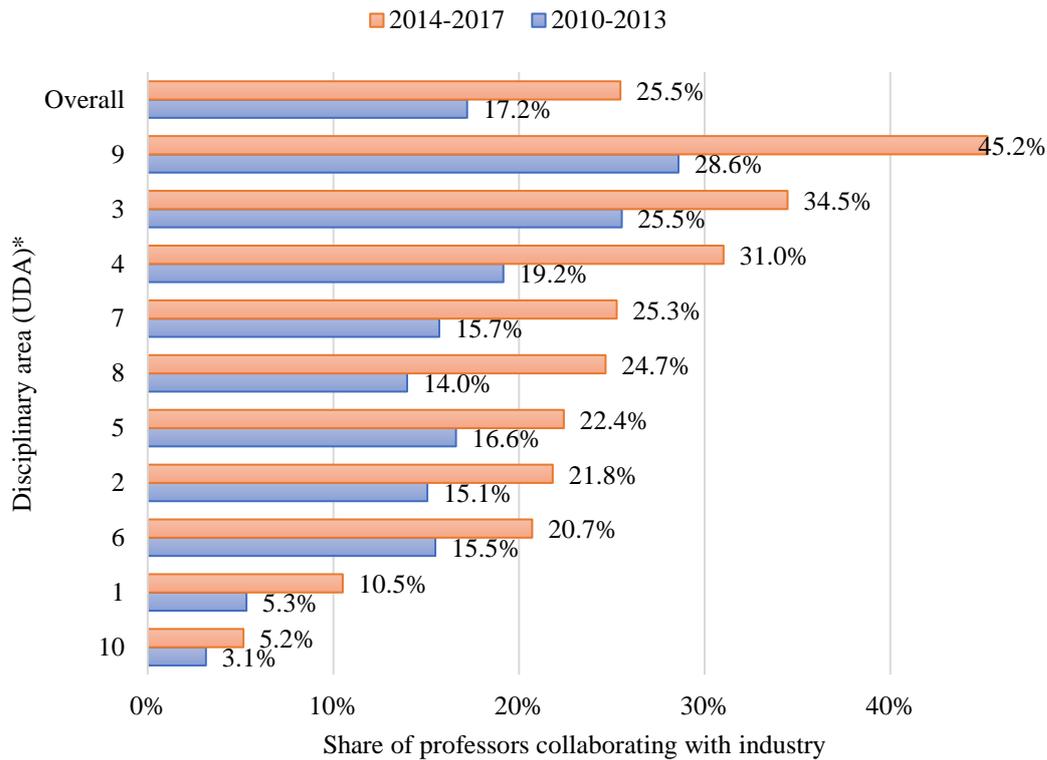

* 1 - Mathematics and computer science, 2 - Physics, 3 - Chemistry, 4 - Earth sciences, 5 - Biology, 6 - Medicine, 7 - Agricultural and veterinary sciences, 8 - Civil engineering, 9 - Industrial and information engineering, 10 – Psychology

Table 7 reports the variation in impact of publications observed between the two periods, in each UDA. Columns 2 and 3 indicate that the difference is negative in all except Mathematics and computer science (UDA 1), and Physics (UDA 2). The decrease in average impact of publications in the different disciplines is mainly due to professors who collaborate with industry across both periods (A ∩ B); i.e. what was observed at the overall level (Table 3) seems confirmed for individual UDAs. Observing turnover of faculty in the disciplines, the young researchers entering the system record a higher impact of publications from private-company collaboration (B - A) than did the older, leaving colleagues (A - B): this holds true in all UDAs but Mathematics and computer science.

*Table 7. Average impact of publications resulting from academic-industry collaboration, by UDA*

| UDA | A 2010-13 | B 2014-17 | Variation | A ∩ B 2010-13 | A ∩ B 2014-17 | Variation | A - B 2010-13 | B - A 2014-17 | Variation |
|---|---|---|---|---|---|---|---|---|---|
| 1 | 0.713 | 1.077 | 0.363 | 0.687 | 1.186 | 0.499 | 0.963 | 0.593 | -0.370 |
| 2 | 0.929 | 1.560 | 0.631 | 0.939 | 1.418 | 0.479 | 0.835 | 2.771 | 1.936 |
| 3 | 0.865 | 0.806 | -0.059 | 0.864 | 0.792 | -0.072 | 0.882 | 0.959 | 0.077 |
| 4 | 0.861 | 0.851 | -0.009 | 0.868 | 0.853 | -0.016 | 0.787 | 0.829 | 0.042 |
| 5 | 1.083 | 0.950 | -0.134 | 1.124 | 0.951 | -0.173 | 0.750 | 0.940 | 0.190 |
| 6 | 1.155 | 0.995 | -0.160 | 1.182 | 1.003 | -0.179 | 0.891 | 0.923 | 0.032 |
| 7 | 1.026 | 0.927 | -0.099 | 1.056 | 0.932 | -0.124 | 0.668 | 0.878 | 0.209 |
| 8 | 0.889 | 0.640 | -0.248 | 0.901 | 0.630 | -0.271 | 0.667 | 0.719 | 0.052 |
| 9 | 0.939 | 0.776 | -0.163 | 0.948 | 0.766 | -0.183 | 0.753 | 0.863 | 0.110 |
| 11 | 0.828 | 0.617 | -0.211 | 0.846 | 0.697 | -0.149 | 0.119 | 0.363 | 0.244 |

* "A" = professors in the 2010-2013 dataset; "B" = professors in the 2014-2017 dataset

With minor variations, we repeated the logit regression above (overall level) on the individual UDAs, for detection of the weight of individual and contextual drivers on collaboration with industry. The stratification of the dataset by UDA reduces the number of observations available for modeling, therefore, to maintain significance of results:
- for age, only two dummies were considered, relative to three levels (below 46 years old as baseline, 46-60, and over 60);
- for university location, only two dummies were considered relative to three localizations (south as baseline, center, and north).

For reasons of space, the regression results are shown in three tables.

Table 8 reports the odds ratios on some professor personal traits. At the UDA level, the effect of gender is almost never statistically significant: the odds ratio is often greater than unity, specifically in Chemistry, Earth sciences, Medicine and Industrial and information engineering (UDAs 3, 4, 6 and 8). Medicine, the discipline with largest number of professors, is obviously the one with the greatest number of significant effects, which are just as obviously in agreement with the observations at overall level: propensity to industry collaboration increases with academic rank but decreases considerably with age, always with significant effect, and for this discipline, always with the effects from both personal traits systematically greater in the first period: i.e. rank and age seem less influential with time. In the other ADUs, rank and age effects are generally nonsignificant, with the exception of the specific case of age over 60. In the UDAs in which odds ratios are statistically significant in both periods, there is a slight reduction in the gap between young colleagues (less than 46) and the over 60s in terms of propensity, with the exception of ADU 8 (Civil engineering). Concerning the trait of academic rank, all other conditions equal, Physics (UDA 2) presents the remarkable case of the difference in propensity to collaboration of full professors vs assistants increasing from +60% in the first period to +125% in the second.

*Table 8: Italian professors, logit regression by UDA: Odds ratios of personal traits as drivers of propensity to collaborate with industry*

|  | Gender | | Age: 46-60 | | Age: Over 60 | | Rank: Associate | | Rank: Full | |
| --- | --- | --- | --- | --- | --- | --- | --- | --- | --- | --- |
| UDA | 2010-13 | 14-17 | 10-13 | 14-17 | 10-13 | 14-17 | 10-13 | 14-17 | 10-13 | 14-17 |
| 1 | 0.697 | 0.568*** | 1.310 | 0.832 | 1.038 | 0.496** | 0.772 | 0.964 | 0.866 | 1.139 |
| 2 | 0.891 | 0.849 | 0.847 | 0.772* | 0.422*** | 0.488*** | 0.957 | 1.304* | 1.600** | 2.248*** |
| 3 | 1.005 | 1.025 | 1.097 | 0.900 | 0.615** | 0.616** | 0.839 | 0.948 | 0.801 | 1.195 |
| 4 | 1.049 | 1.054 | 0.856 | 1.172 | 0.330** | 1.314 | 1.135 | 0.962 | 1.597 | 0.903 |
| 5 | 0.964 | 0.920 | 1.190 | 1.140 | 0.933 | 0.818 | 0.986 | 1.091 | 1.091 | 1.058 |
| 6 | 1.148* | 1.065 | 0.715*** | 0.782*** | 0.615*** | 0.631*** | 1.462*** | 1.441*** | 1.724*** | 1.581*** |
| 7 | 0.813 | 0.928 | 1.079 | 0.848 | 1.343 | 0.823 | 1.123 | 1.003 | 0.772 | 1.069 |
| 8 | 1.280 | 0.952 | 0.907 | 0.650** | 0.416** | 0.407*** | 1.030 | 1.256 | 1.090 | 1.756** |
| 9 | 0.928 | 0.912 | 0.882 | 0.813** | 0.620*** | 0.667*** | 1.314*** | 1.077 | 1.312** | 1.134 |
| 10 | 0.917 | 0.613* | 1.041 | 0.527* | 0.586 | 0.313* | 1.628 | 2.682*** | 1.402 | 1.874 |

*Statistical significance: *p-value <0.10, **p-value <0.05, ***p-value <0.01*

Table 9 reports the odds ratios of the covariates related to the academics' scientific activity. Columns 2 and 3 refer to the effect of the professor's overall productivity: where significant, this always has a positive sign (odds ratio greater than one), however the magnitude of this effect becomes less over time in all disciplines except Earth science (UDA 4). Contrastingly, the weight of the diversification/specialization of a faculty member's scientific portfolio on their propensity to collaboration increases: for "high diversification" the increase is seen in six out of ten UDAs, while for the dual variable "high specialization" the increase is in the two largest disciplines (Biology, Medicine), and in Physics.

*Table 9: Italian professors, logit regression by UDA: Odds ratios of personal scientific activity traits as drivers of propensity to collaborate with industry*

|     | Output    |          | Highly diversified |          | Highly specialized |          |
| --- | --------- | -------- | ------------------ | -------- | ------------------ | -------- |
| UDA | 2010-13   | 2014-17  | 2010-13            | 2014-17  | 2010-13            | 2014-17  |
| 1   | 1.067***  | 1.013*** | 1.211              | 1.777*** | 0.259***           | 0.304*** |
| 2   | 0.999     | 0.999    | 1.611***           | 1.744*** | 0.587***           | 0.388*** |
| 3   | 1.042***  | 1.032*** | 1.218              | 1.463*** | 0.653***           | 0.699**  |
| 4   | 1.051***  | 1.064*** | 1.162              | 1.767**  | 0.546***           | 0.999    |
| 5   | 1.044***  | 1.038*** | 1.157              | 1.272*** | 0.583***           | 0.366*** |
| 6   | 1.035***  | 1.030*** | 1.267***           | 1.294*** | 0.631***           | 0.554*** |
| 7   | 1.084***  | 1.058*** | 1.323              | 1.168    | 0.662***           | 1.024    |
| 8   | 1.046***  | 1.040*** | 1.247              | 0.967    | 0.699*             | 0.969    |
| 9   | 1.049***  | 1.035*** | 1.190              | 1.040    | 0.845**            | 0.989    |
| 10  | 1.049***  | 1.036*** | 1.599              | 1.559    | 0.261**            | 0.290**  |

*Statistical significance: \*p-value <0.10, \*\*p-value <0.05, \*\*\*p-value <0.01*

Finally, Table 10 shows the breakdown by UDA of odds ratios related to some contextual covariates. In both periods, the effect of the peers variable is very strong in all UDAs, lacking significance only in Psychology, 2010-2013. The value of the effect, however, becomes lower in 2014-2017, in all disciplines but Earth science, and Civil engineering (UDAs 4 and 8). The effect of the size of the professor's disciplinary context (columns 4 to 7) is less clear: the odds ratios for propensity to industry collaboration often lack statistical significance. Medicine (UDA 6) is usually the discipline with the clearest regression results: these indicate an increase over time in the negative correlation between the covariate size (number of colleagues in same discipline, same university) and the response variable. Compared with physicians operating in a "small" setting:

- colleagues operating in a medium-sized context tend less to collaborate with industry (-26.5% in the first period, odds ratio 0.735; -45.6% in the second, odds ratio 0.544).
- colleagues operating in a "large" context are the least disposed to collaborate with industry (-35.8% in first period, -49.1% in second).

In Table 10, the last two columns show the variation in the weight of geographic location on the academic's propensity to collaborate. Taking "South" as baseline, we note that the probability of an academic to engage in collaborations with industry does not increase systematically with latitude. In Mathematics (UDA 1), for a professor from the center compared to one from the south, this probability is 41.4% lower in the first period (odds ratio 0.586), and 35.6% lower in the second (odds ratio 0.644). A similar pattern can be observed when considering the North, although the effects are not significant.

In general, between the first and second period, the odds ratios increase, i.e. the weight of the geographic factor increases. The exceptions are Biology (UDA 5), and Industrial and information engineering (UDA 9) where the negative disparity in propensity of professors from the South with respect to the North diminishes from the first to the second period, and in Medicine, where the disparity recorded in the first period becomes not significant in the second.

*Table 10: Italian professors: Odds ratios of contextual covariates of the propensity to collaborate with industry. (Logit regression, dependent variable: 1 in case of publications in co-authorship with industry, 0, otherwise.)*

|     | Peers    |          | Univ size: Medium |         | Univ size: Large |          | Univ area: Center |          | Univ area: North |          |
| --- | -------- | -------- | ----------------- | ------- | ---------------- | -------- | ----------------- | -------- | ---------------- | -------- |
| UDA | 2010-13  | 2014-17  | 10-13             | 14-17   | 10-13            | 14-17    | 10-13             | 14-17    | 10-13            | 14-17    |
| 1   | 4.606*** | 3.665*** | 0.772             | 0.563*  | 0.673            | 0.620    | 0.586*            | 0.644**  | 0.745            | 0.939    |
| 2   | 3.162*** | 2.398*** | 0.542**           | 1.455   | 0.445***         | 1.375    | 0.690**           | 1.027    | 0.812            | 1.056    |
| 3   | 3.913*** | 3.511*** | 0.896             | 0.846   | 0.653**          | 0.596**  | 1.277*            | 1.350**  | 1.405***         | 1.503*** |
| 4   | 3.263*** | 3.470*** | 0.655             | 0.527*  | 0.795            | 0.465**  | 1.020             | 1.125    | 1.096            | 1.202    |
| 5   | 2.861*** | 2.094*** | 0.846             | 1.126   | 0.521***         | 0.848    | 1.183             | 1.024    | 1.268**          | 1.195*   |
| 6   | 3.204*** | 2.358*** | 0.735*            | 0.544** | 0.642***         | 0.509*** | 0.986             | 1.014    | 1.243***         | 1.078    |
| 7   | 3.756*** | 2.745*** | 1.838**           | 1.543** | 1.145            | 1.137    | 1.001             | 2.049*** | 1.385**          | 2.059*** |
| 8 | 3.293*** | 3.719*** | 0.559* | 0.579** | 0.566* | 0.587** | 0.920 | 1.782*** | 1.372 | 1.958*** |
| 9 | 3.572*** | 3.107*** | 0.789 | 1.088 | 0.714** | 0.973 | 1.110 | 1.283*** | 1.282*** | 1.196** |
| 10 | 1.251 | 2.369*** | 0.876 | 0.492 | 0.638 | 1.468 | 1.041 | 0.394* | 0.637 | 0.664 |

*Statistical significance: \*p-value <0.10, \*\*p-value <0.05, \*\*\*p-value <0.01*

## 5. Conclusions

In the current so called knowledge-based economy, knowledge is increasingly becoming a distinctive competence to achieve competitive advantage, and to sustain the economic growth of nations. For this reason, universities, as the loci of knowledge creation, not only accept the responsibility of technology transfer as their third mission, alongside education and research (Etzkowitz & Leydesdorff, 1995), but also face increasing pressure to devote particular attention to this area (Todorovic, McNaughton, & Guild, 2011; Etzkowitz & Leydesdorff, 1995; Etzkowitz, 1983). The "public-private transfer" is generally conceived of as research at universities and public institutions, with the results then exploited by industry. Companies, however, can gain new knowledge more effectively and quickly by co-creating the new knowledge through the activities of joint public-private research projects. For this reason, various countries have developed incentivizing schemes for academic-industry collaborations (Davenport, Davies, & Grimes, 1998; Debackere & Veugelers, 2005).

In formulating and targeting such incentive systems, it is critical to understand the drivers underlying the academics' propensity to engage in collaboration with industry. When setting reward systems and performance indicators, policy makers and administrators must be aware of the relative weight of the drivers, and how these vary over time, in order to successfully design and evaluate the effectiveness of their initiatives, as well as readjust them.

Incentive systems can be confounded by inherent disincentives, and even by policies operating at cross-purposes, meaning that policy-makers and administrators require a complete vision of the drivers at play in the design and operation of strategies for public-private collaboration. First of all, the university-industry partnership will face transactions costs (Belkhodja & Landry, 2005; Drejer & Jørgensen, 2005), growing with the cultural and cognitive distance of the team members (Abramo, D'Angelo, Di Costa, & Solazzi, 2011). Another potential deterrent is that academics, increasingly subject to evaluation of their scientific activity, might prefer other personal research strategies with apparently greater chance of favorable evaluation. It is known, for example, that on average, intra-university co-authored publications have higher impact than public-private co-authored works, and compose a higher share of highly-cited publications (Abramo, D'Angelo, & Di Costa, 2020): these sorts of facts, and their awareness, could lead to difficult choices at different levels.

In this study, we show that the share of Italian professors collaborating with industry rose from 17.2 percent in 2010-2013 to 25.5 percent in 2014-2017. The explanation for this lies partly in the increasing propensity of faculty incumbents to collaborate with industry, and partly in the arrival of new hires with higher propensity to engage, compared to the older faculty departing in retirement. A somewhat pessimistic view would be that the increasing resort to private collaboration descends solely from tough economic times, budget constraints, and the progressive thinning of resources available to research institutions and professors. But we could also expect that both of these phenomena would be due to administrators and professors gradually accepting the importance of the university's third mission, responding to pressure and reward to open up their research agendas to the needs of the local and national production system. Although the Ministry of Universities and Research has itself not yet installed specific incentives for universities in this sense, within its performance-based research funding scheme, the universities' entrepreneurial and technology transfer activities are now being assessed for the first time in the current national evaluation exercise.

The type profile of the professor collaborating with industry remains stable over the years analyzed: the typical academic is a male, full professor, relatively young, with highly diversified research activity, operating within a team of disciplinary colleagues, themselves with a high propensity to collaboration, on staff in a public university in northwestern Italy.

The importance of the drivers of academic engagement in collaboration, relative to one another, remains quite stable over the two periods examined. The variations of marginal effects from personal traits, such as gender and age, are modest. Most noticeable are the decrease in marginal effect of the individual's academic rank over time, and the increasing effect of research diversification.

The contextual driver with the greatest weight on likelihood of collaborating with industry is the presence of peers who themselves share a high propensity to collaboration. However, this effect loses weight over the two periods examined, as does the effect of smaller university sizes. The effect of localization in the more northern, industrial regions of Italy, instead gains weight.

Aligned with Tijssen (2012), also in the Italian case the propensity to collaborate with industry varies across disciplines. The variations observed at overall level are confirmed as similar at discipline level, with very few exceptions.

Any bibliometric approach to the analysis of academic-industry collaborations will be subject to intrinsic limits, among these: not all research collaborations lead to an indexed publication, and not all joint co-authored publications will reflect a real public-private collaboration. For this reason, we recommend caution in interpreting the results.

A useful follow-on to the current research could be to extend the analysis to previous periods, using panel data. Another direction could be to compare the effects of geographic proximity on the intensity of collaborations, looking at both academic-industry and intra-sector partnerships. Finally, the methodology can be easily applied to other nations, whereby natives can easily distinguish and reconcile public and private affiliations. This would allow cross-country comparisons for a better understanding of the phenomenon.